\documentclass{article}
\usepackage{spconf,amsmath,graphicx}
\usepackage{mathrsfs}
\usepackage{verbatim}
\usepackage{array}
\usepackage{bm}
\usepackage{caption}
\usepackage{diagbox}
\usepackage{amssymb}
\usepackage{enumitem}
\setlist{nosep, leftmargin=14pt}
\usepackage{mwe} 

\title{Local implicit neural representations for multi-sequence MRI translation}
\name{Yunjie Chen$^{\star}$, Marius Staring$^{\star}$, Jelmer M. Wolterink$^{\dagger}$, Qian Tao$^{\star,\ddag}$}
\address{$^{\star}$Department of Radiology, Leiden University Medical Center, the Netherlands\\
        $^{\dagger}$Department of Applied Mathematics, Technical Medical Center, University of Twente, the Netherlands\\
        $^{\ddag}$Department of Imaging Physics, Delft University of Technology, the Netherlands}
\begin{document}
%
\maketitle
\begin{abstract}
In radiological practice, multi-sequence MRI is routinely acquired to characterize anatomy and tissue. However, due to the heterogeneity of imaging protocols and contra-indications to contrast agents, some MRI sequences, e.g. contrast-enhanced T1-weighted image (T1ce), may not be acquired. This creates difficulties for large-scale clinical studies for which heterogeneous datasets are aggregated. Modern deep learning techniques have demonstrated the capability of synthesizing missing sequences from existing sequences, through learning from an extensive multi-sequence MRI dataset. In this paper, we propose a novel MR image translation solution based on \emph{local implicit neural representations}. We split the available MRI sequences into local patches and assign to each patch a local multi-layer perceptron (MLP) that represents a patch in the T1ce. The parameters of these local MLPs are generated by a hypernetwork based on image features. Experimental results and ablation studies on the BraTS challenge dataset showed that the local MLPs are critical for recovering fine image and tumor details, as they allow for local specialization that is highly important for accurate image translation. Compared to a classical pix2pix model, the proposed method demonstrated visual improvement and significantly improved quantitative scores (MSE \(0.86 \times 10^{-3}\) vs. \(1.02 \times 10^{-3}\) and SSIM 94.9 vs 94.3).
\end{abstract}

\begin{figure*}[tb]
 \centering
\includegraphics[width=\linewidth]{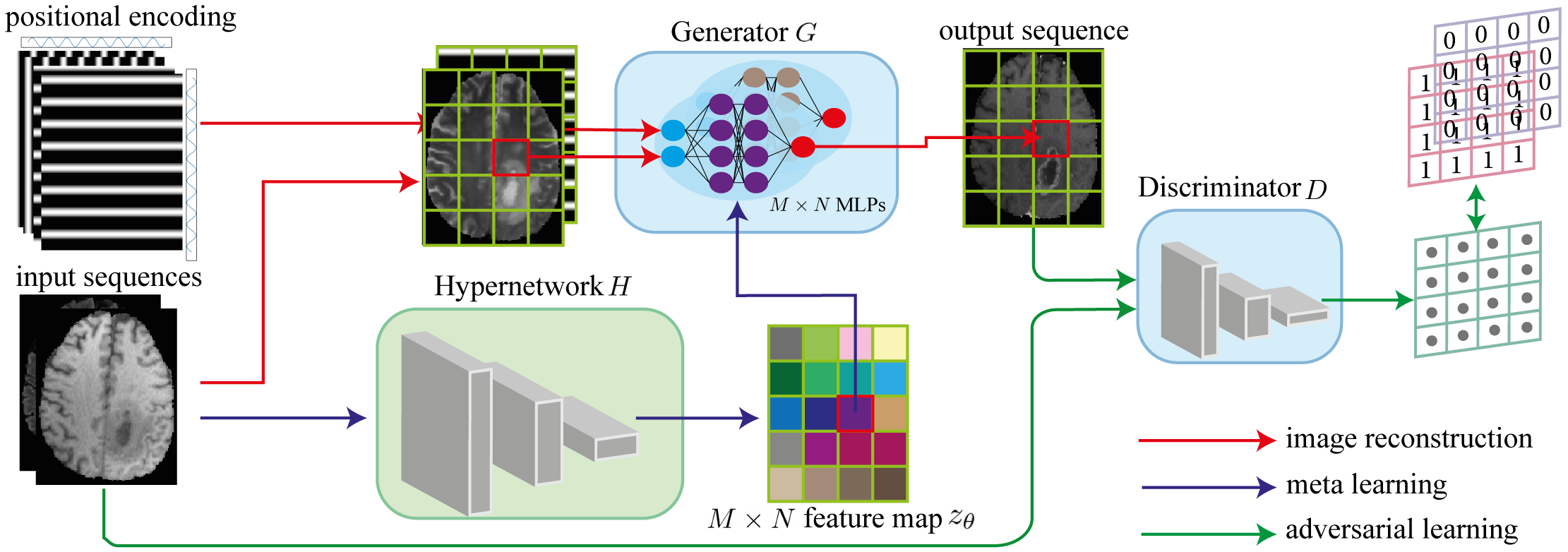}
\caption{Illustration of our method. The input image and encoded coordinates are split into patches.\(M \times N\) MLPs, which use the output of a hypernetwork as weights, are assigned to each image patch and output a corresponding patch in the target domain. During training, we use a patch discriminator for adversarial learning to improve translation quality.}
\label{fig:architecture}
\end{figure*}

\section{Introduction}
\label{sec:intro}

Magnetic resonance imaging (MRI) is one of the most important image modalities in clinical radiology because of its excellent soft tissue contrast and versatility. One distinguishing advantage of MRI is the availability of various sequences that provide complementary information to help clinicians characterize the anatomy in a comprehensive way. Recent research on MRI-based regression and segmentation also demonstrated the value of multi-sequence MRI in machine learning studies \cite{Cherubini2016}. However, due to the heterogeneity of imaging protocols and contra-indications of contrast agents, some MRI sequences are not always acquired, especially those requiring contrast administration. This may influence the usability of machine learning models originally trained on multi-sequence MRI.

A number of methods have been developed to address this missing sequence problem \cite{Azad2022}. The basic idea of earlier attempts is to build a common latent space shared by all modalities such as Hetero-Modal Image Segmentation (HeMIS) \cite{Havaei2016}. Such methods tend to focus on specific tasks like regression or segmentation and offer accurate results without explicitly generating the missing images, which however damages the reliability and interpretability of the improvement achieved. Nowadays, generative adversarial networks (GANs) are commonly used to directly synthesize missing modalities or MRI sequences from existing images. For instance, Sharma \cite{Sharma2020} proposed a Multi-Modal Generative Adversarial Network (MM-GAN) which accepts arbitrary MRI sequences as input to generate missing sequences. In spite of the promising results these studies have shown, the synthesis of new information is still less than satisfactory, especially when synthesizing enhanced lesion regions from non-enhanced images \cite{Azad2022}.

To address these limitations, we introduce implicit neural representations (INRs) as the generator in a GAN framework for multi-sequence MRI translation \cite{Xie2022}. INRs is a multi-layer perceptron (MLP) that takes as input a continuous coordinate and provides an estimation of a quantity at that point as output. In addition, we propose to encode prior knowledge from the dataset using a hypernetwork which can enhance generalization and performance \cite{Jayakumar2020}. The main contributions of this work are as follows: 1) We propose a GAN-based model using INRs as generator which includes meta-learning and locally specific MLPs to synthesize a T1ce from a T2-weighted image (T2), T1-weighted image (T1) and Fluid attenuation inversion recovery (Flair); 2) We show that our method improves translation performance in the BraTS challenge dataset both qualitatively and quantitatively compared to a traditional GAN which uses a U-Net shaped generator; 3) We design an ablation study and further demonstrate the role of the locally specific MLPs for MRI translation.

\section{Methods}
\label{sec:method}
\subsection{Generative adversarial network}
\label{sec:gan}
For simplicity, we formulate our method with single input and single-output, but our method can be extended to a multi-input setting by concatenating multiple sequences as a multi-channel input. The goal is to learn a non-linear mapping between two image domains. Given a set of paired images: T2 (domain \(\mathcal{S}\)) and T1ce (domain \(\mathcal{T}\)), we aim to learn the mapping \(G: s \rightarrow t\), with \(s \in \mathcal{S}\) and \(t \in \mathcal{T}\). GAN-based models like pix2pix \cite{Isola2017}\cite{Zhu2017} are widely used for this:
\begin{align}
\begin{split}
L_{\mathrm{GAN}}(G,D) &= \mathbb{E}_{s,t}[\log D(s,t)] \\
             &+ \mathbb{E}_{s,t}\log(1-D(s,G(s,\bm{z}))),\label{eq:1}
\end{split}
\end{align}
where \(G\) is the generator, \(D\) is the discriminator, and \(\bm{z}\) is the latent code which is usually sampled from a prior Gaussian distribution. In a pix2pix model, a U-Net is used as a generator, often with a patch discriminator (PatchGAN) \cite{Isola2017}. In the following sections, we introduce the main idea of implicit neural representations and how this technique can be applied to GAN-based multi-sequence MRI translation.

\subsection{Implicit neural representations}
\label{sec:inr}
Recent research on INRs has shown that complex real-world signals can be represented as a continuous function by a fully connected network \cite{Mildenhall2021}. In our case, we use an MLP to represent the intensity values of a T1ce. The MLP can be conditioned, such that the function that it represents depends on other information. Here, we condition the MLP on the source image intensity at its input coordinate. Then for each pixel, the image translation mapping function \(G\) can be cast into a pixel-wise mapping as follows: \(G(\bm{x},s_{\bm{x}})=t_{\bm{x}}^{'}\), where \(\bm{x}=(x,y) \in \mathbb{R}^2\) is the 2D coordinate, \(s_{\bm{x}} \in \mathbb{R}\) is the pixel value in the input image and \(t_{\bm{x}}^{'} \in \mathbb{R}\) is the pixel value in the output image. In addition, we introduce a latent code \(\bm{z}\) which encodes prior knowledge which is not present in the input image, e.g. contrast-enhanced information of the tumor. To do this, we use a convolutional neural network \(H\) as a hypernetwork with trainable parameters \(\theta\) which takes the source image \(s\) as input, to output the weights used by the MLP. As a result, the mapping function will be conditioned on every different input instance: 
\begin{align}
G(\bm{x},s_{\bm{x}};\bm{z}_{\theta}(s)) = t_{\bm{x}}^{'}.
\end{align}
The overall pipeline is shown in Fig.~\ref{fig:architecture}.

\subsection{Locally specific MLPs}
\label{sec:mlp}

To achieve a focused local representation, we adopt the idea of multiple locally specific MLPs from ASAP-Net \cite{Shaham2021}, instead of using a single MLP to represent the entire image. As shown in Fig.~\ref{fig:architecture}, the input image and coordinates are split into \(M \times N\) patches and each patch is assigned an unique MLP. The hypernetwork \(H\) downsamples the input image to the same resolution and outputs weights used by these MLPs. This enables these MLPs to be conditioned on the whole input image while still being able to encapsulate unique information from the local region.

Since an MLP with ReLU activation functions that directly operates on Cartesian coordinates has limited representation power due to a spectral bias, we employ a positional encoding \(\gamma: \mathbb{R}^2 \rightarrow \mathbb{R}^M\) to map the coordinates into a higher dimensional space~\cite{Mildenhall2021}. Unlike ASAP-net, in which coordinates have the same positional encoding in each patch, we first globally normalize the coordinates into $[0,1]$, and then apply sinusoids to each of the two normalized coordinates separately: $\gamma(\bm{x})=[\sin(2^0\pi \bm{x}), \cos(2^0\pi \bm{x}), \ldots,\allowbreak \sin(2^{i-1}\pi \bm{x}), \cos(2^{i-1}\pi \bm{x})]$, with $i$ being the frequency parameter dependant on the task. This provides the network with the ability to fit high-frequency signals, while also giving every position in the image a unique representation.

\subsection{Objective}
\label{sec:objective}
In addition to implicit neural representation, we employ an adversarial learning strategy to further improve the quality of the synthesized image \cite{Isola2017}. The hypernetwork and MLPs are jointly optimized using the GAN loss (\ref{eq:1}). For an image translation task, the output of the generator should be as close as possible to the target image. We use the \(\ell_1\) loss as the reconstruction loss, which can produce sharper results compared to mean squared error (MSE): $L_{\mathrm{rec}}(G)=\|t^{'}-t\|_{1}=\|G(s;\bm{z}_\theta(s))-t\|_{1}$. The final objective function is then:
\begin{align}
\theta = \arg \min\limits_{\theta} \max\limits_D L_{\mathrm{GAN}}(G,D)+\lambda_{\mathrm{rec}} L_{\mathrm{rec}}(G).
\end{align}
For more stable training and faster convergence, we add Gaussian noise to the real and fake images before we pass them to the discriminator \cite{Martin2017}.

\begin{figure}[tb]
 \centering
\includegraphics[width=\columnwidth]{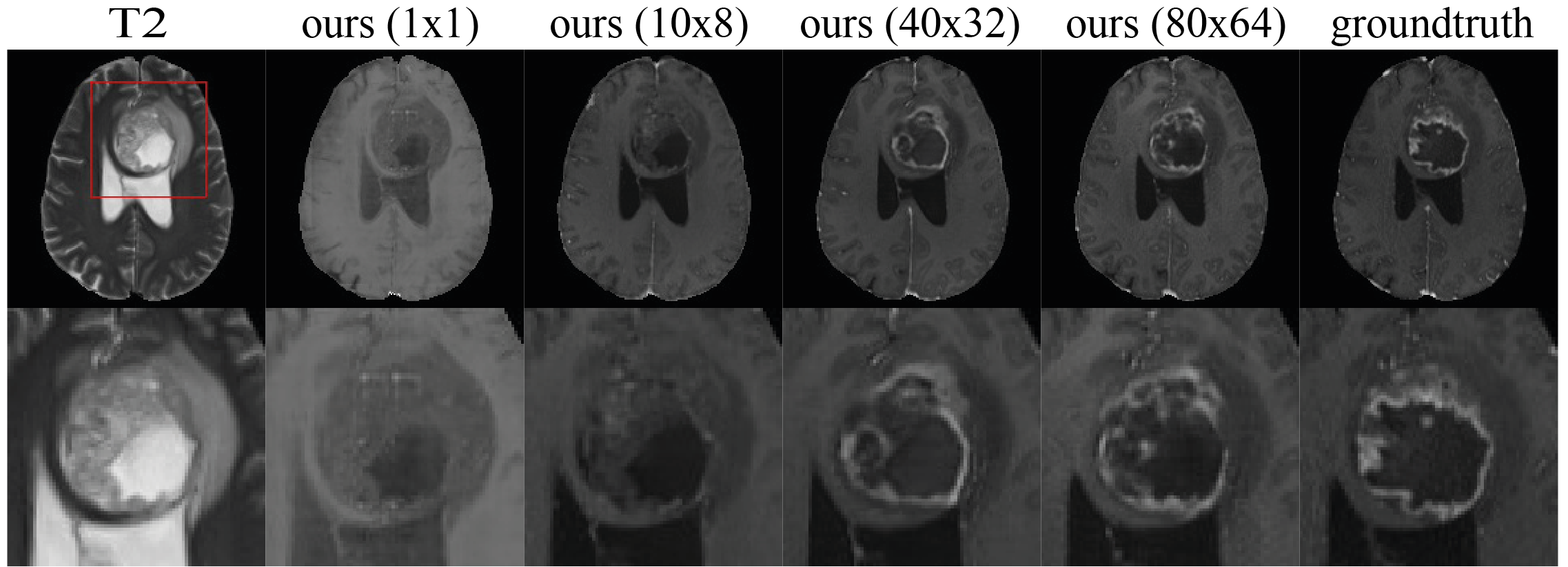}
\caption{The influence of choices of \(M \times N\). Whole brain MRI are shown in the top row and corresponding zoomed-in results are shown in the bottom row.}
\label{fig:patchcompare}
\end{figure}

\section{Experiments and Results}
\label{sec:rst}
\subsection{Dataset}
\label{ssec:dataset}
We evaluated our methods on the BraTS2018 challenge dataset~\cite{Menze2015}, which provides co-registered MRI sequences at 1 mm\(^3\) isotropic resolution. Each patient has four sequences: T1, T2, T1ce, and Flair. We randomly selected 296 patients for training and 76 patients for testing. Before training, all image intensities are normalized to [-1,1] and non-zero cropping is performed to reduce the size of images.

\subsection{Experiments}
\label{ssec:experiments}
In our experiment, we synthesize T1ce using two input settings: single-input (T2) and multi-input (T1, T2, Flair). We compare the performance of our method with pix2pix \cite{Isola2017}, with the only difference being the generator. The discriminator consists of three \(4 \times 4\) convolutional blocks with stride 2 and two \(4 \times 4\) convolutional blocks with stride 1. The hypernetwork contains a downsampling module, seven \(3 \times 3\) convolutional blocks, and a final convolutional layer to output the desired number of parameters of the MLPs. For a fair comparison, we used an 8-layer MLP (100K parameters) for models with only one global MLP, and a 5-layer MLP (18K parameters) for models with locally specific MLPs. So in total, our models contain 100 million parameters (hypernetwork and MLPs). The models were optimized using Adam with an initial learning rate of \(10^{-4}\) with a step decay schedule for 60 epochs. We set the batch size to 8, \(\lambda_{\mathrm{rec}}\) to 100 and $i$ to 6 for postional encoding. During training, random flipping is performed and every image is randomly cropped to \(160 \times 128\). We quantitatively evaluate our methods using mean squared error (MSE), structural similarity index (SSIM), and peak signal-to-noise ratio (PSNR). The results of the two translation models are compared statistically using the Wilcoxon signed-rank test.

\begin{table}
\caption{Quantitative T1ce translation results comparing pix2pix with our methods. \(\ast\) indicates significant difference (\(p < .001\)) compared to pix2pix using the same inputs.}
\label{quantitativeresults}
\resizebox{\columnwidth}{!}{\begin{tabular}{l | c | c | c | c }
 \hline
methods & \(M \times N\) & MSE \((\times 10^{-3}) \downarrow\) & SSIM $\uparrow$ & PSNR $\uparrow$ \\ 
 \hline\hline
 pix2pix (m=1) & \diagbox{ }{ } &\(1.02 \pm 1.25\phantom{^*}\) & \(94.3 \pm 1.3\phantom{^*}\) & \(30.98 \pm 2.58\phantom{^*}\) \\ 
\hline
ours (m=1) & \(1 \times 1\) & \(1.63 \pm 1.42^*\) & \(93.2 \pm 1.5^*\) & \(28.76 \pm 2.50^*\) \\
 ours (m=1) & \(10 \times 8\) & \(1.05 \pm 1.18\phantom{^*}\) & \(94.4 \pm 1.3\) & \(30.80 \pm 2.51\) \\
ours (m=1) & \(20 \times 16\) & \(0.92 \pm 1.20^*\) & \(94.9 \pm 1.2^*\) & \(31.55 \pm 2.71^*\) \\
ours (m=1) & \(40 \times 32\) & \(0.90  \pm 1.11^*\) & \(94.9 \pm 1.2^*\) & \(31.61 \pm 2.79^*\) \\
ours (m=1) & \(\bm{80 \times 64}\) & \(\bm{0.86  \pm 0.91}^*\) & \(\bm{94.9 \pm 1.2}^*\) & \(\bm{31.82 \pm 2.82}^*\) \\
\hline\hline
pix2pix (m=3) & \diagbox{ }{ }& \(0.77 \pm 0.88\phantom{^*}\) & \(95.1 \pm 1.1\phantom{^*}\) & \(32.32 \pm 2.85\phantom{^*}\) \\
\hline
ours (m=3) & \(\bm{40 \times 32}\) & \(\bm{0.74 \pm 1.01}^*\) & \(\bm{95.6 \pm 1.2}^*\) & \(\bm{32.76 \pm 3.15}^*\) \\
\hline
\end{tabular}}
\end{table}

\begin{figure*}[tb]
 \centering
 \includegraphics[width=\linewidth]{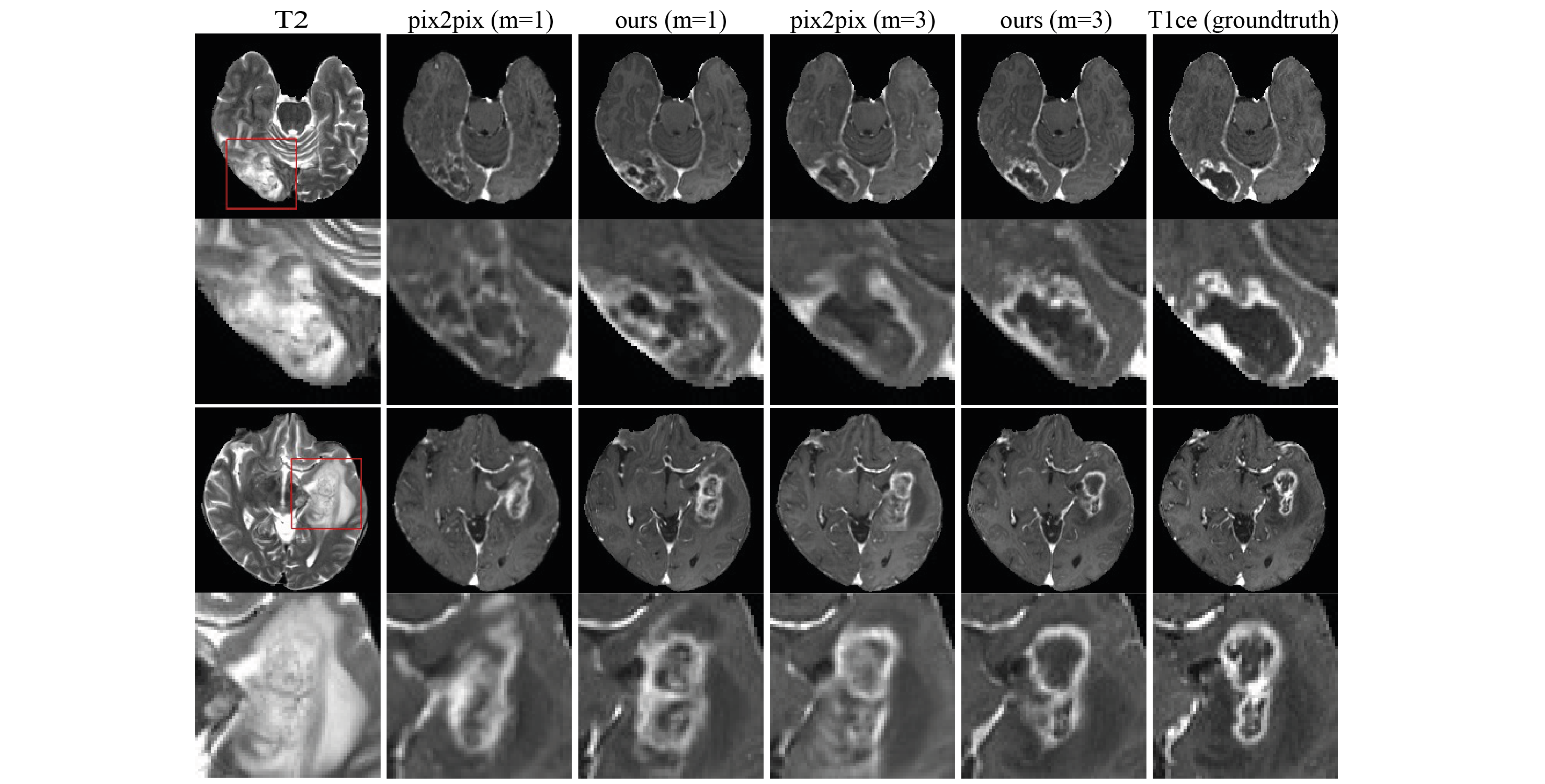}
 \caption{Visual results of pix2pix and our method. Zoomed-in areas are indicated by red rectangles.}
\label{fig:gan_vs_inp}
\end{figure*}

\subsection{MRI translation}
\label{sec:mri translation}
We first compared models with different numbers of MLPs. We set \(M \times N\) to \(1 \times 1\) (one global MLP), \(10 \times 8\), \(20 \times 16\), \(40 \times 32\), \(80 \times 64\) and the results are reported in Table~\ref{quantitativeresults}. The results show that by using multiple locally specific MLPs, we significantly improve the performance. Compared to the pix2pix model, our model gets comparable results when we set \(M \times N\) to \(10 \times 8\). It can be noted that the models uses more MLPs have higher translation quality and we can get best results when \(M \times N\) is set to \(80\times 64\). As shown in Fig.~\ref{fig:patchcompare}, the model with \(40 \times 32\) MLPs and the model with \(80\times 64\) MLPs are almost visually equivalent. Given the week-long training time of the \(80\times 64\) model using Nvidia Tesla V100, we set \(M \times N\) to \(40 \times 32\) for the subsequent experiments.

The overall quantitative results of models using T2 only (\(m=1\)) and using T1, T2, Flair jointly (\(m=3\)) as input are reported in Table~\ref{quantitativeresults}. The results show that our method outperforms pix2pix in all metrics with both input settings. All results between the two models show statistically significant differences \(p < .001\) using the Wilcoxon signed-rank test. As shown in Fig.~\ref{fig:gan_vs_inp}, we can observe that by using INRs, the results are less blurred and exhibit finer details, showing a more obviously enhanced tumor.

\subsection{Locality of MLPs}
\label{sec:specialized MLP}
\begin{figure}[tb]
 \centering
 \includegraphics[width=\columnwidth]{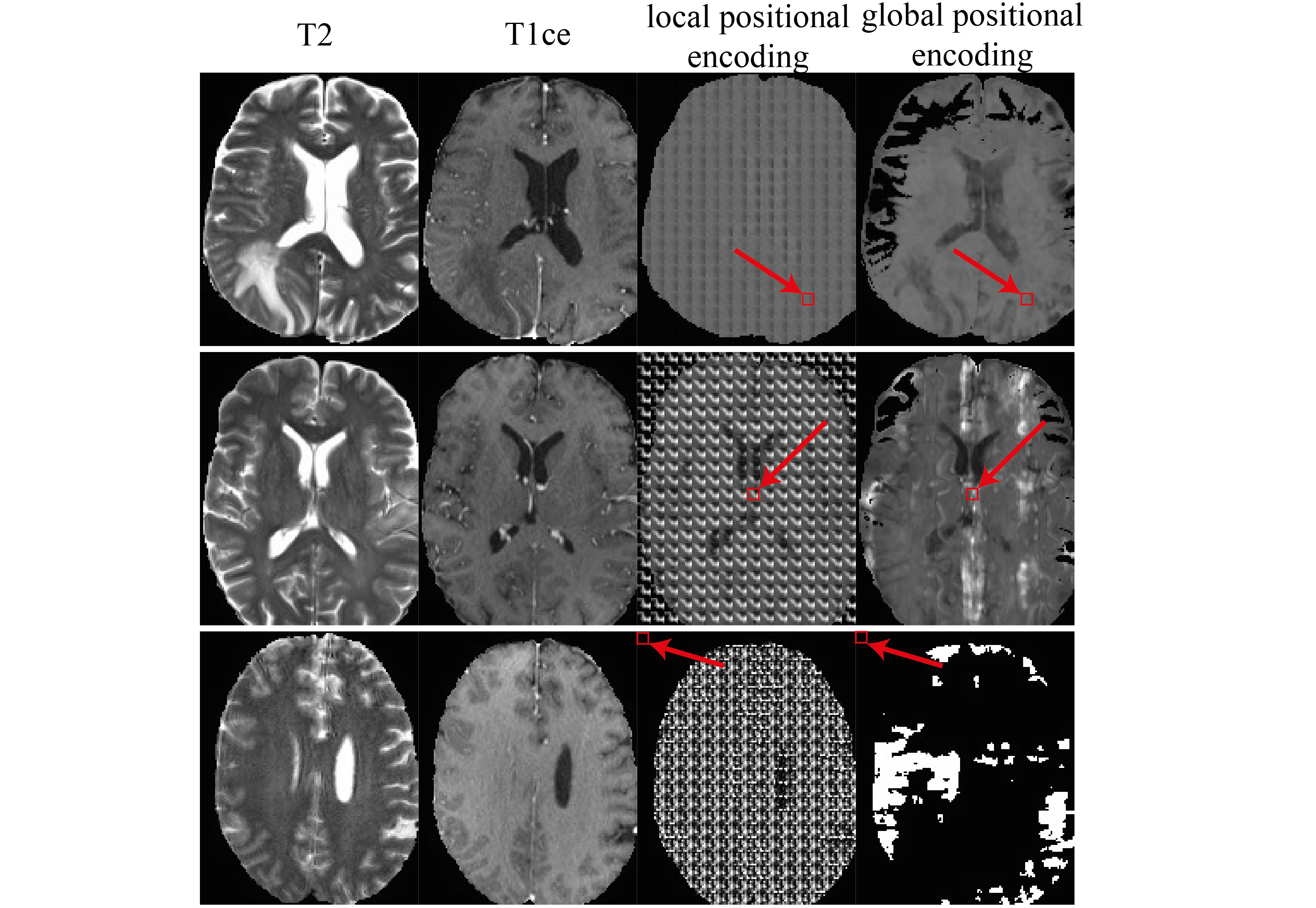}
 \caption{The role of locally specific MLPs: using a single selected MLP, depicted with red rectangle, for generating the entire image.}
\label{fig:patchtest}
\end{figure}

To verify the role of the locally specific MLPs in our model, we performed experiments to forward all pixels through one of the MLPs. As shown in Fig.~\ref{fig:patchtest}, when selecting an MLP inside the white matter (top row), most of the local details of the image are lost. When selecting an MLP located in the cerebrospinal fluid (CSF) of the ventricles (middle row), the locally learned pattern is repeated periodically. When we select a background MLP (bottom row), the prediction plainly failed. This shows that we get specialized MLPs depending on the local image context in our method. On the contrary, the MLPs using local positional encoding tend to output periodic results which indicates the lack of locality.

\section{Discussion and conclusion}
\label{sec:conclusion}
In this paper, we have proposed a novel GAN-based model using implicit neural representations as the generator for multi-sequence MRI translation. Our results on the BraTS dataset show that our methods improved the quality of synthesized T1ce both visually and quantitatively. Compared to the pix2pix method which uses a U-Net shaped generator, our results are overall sharper and recover more details in the enhanced tumor region. We designed an ablation study that demonstrated the specific function of locally specific MLPs in the model, which appeared critical for restoring image detail. Future extensions of this work could be leveraging the synthesized image in a downstream task like segmentation. Furthermore, by using INRs we get a unique representation for each input data, which makes it promising to introduce prior knowledge from follow-up scans in a longitudinal study. In summary, our preliminary experiments have shown promising results in using INRs in MRI translation.
\section{Compliance with Ethical Standards}
\label{sec:Ethical}
The research was conducted using human subject data made available in open access. Ethical approval was not required as confirmed by the license attached with the open access data.

\bibliographystyle{IEEEbib}
\bibliography{references}

\begin{thebibliography}{10}

\bibitem{Cherubini2016}
A.~Cherubini, M.~E. Caligiuri, P.~Péran, U.~Sabatini, C.~Cosentino, and
  F.~Amato,
\newblock ``Importance of multimodal {MRI} in characterizing brain tissue and
  its potential application for individual age prediction,''
\newblock {\em IEEE Journal of Biomedical and Health Informatics}, vol. 20, no.
  5, pp. 1232--1239, 2016.

\bibitem{Azad2022}
R.~Azad, N.~Khosravi, M.~Dehghanmanshadi, J.~Cohen-Adad, and D.~Merhof,
\newblock ``Medical image segmentation on {MRI} images with missing modalities:
  A review,''
\newblock {\em arXiv preprint arxiv:2203.06217}, 2022.

\bibitem{Havaei2016}
M.~Havaei, N.~Guizard, N.~Chapados, and Y.~Bengio,
\newblock ``{HeMIS}: Hetero-modal image segmentation,''
\newblock in {\em International Conference on Medical Image Computing and
  Computer-Assisted Intervention}, 2016, pp. 469--477.

\bibitem{Sharma2020}
A.~Sharma and G.~Hamarneh,
\newblock ``Missing {MRI} pulse sequence synthesis using multi-modal generative
  adversarial network,''
\newblock {\em IEEE Transactions on Medical Imaging}, vol. 39, no. 4, pp.
  1170--1183, 2020.

\bibitem{Xie2022}
Y.~Xie, T.~Takikawa, S.~Saito, O.~Litany, S.~Yan, N.~Khan, F.~Tombari,
  J.~Tompkin, V.~sitzmann, and S.~Sridhar,
\newblock ``Neural fields in visual computing and beyond,''
\newblock {\em Computer Graphics Forum}, vol. 41, no. 2, pp. 641--676, 2022.

\bibitem{Jayakumar2020}
S.~M. Jayakumar, W.~M. Czarnecki, J.~Menick, J.n Schwarz, J.~Rae, S.~Osindero,
  Y.~W. Teh, T.~Harley, and R.~Pascanu,
\newblock ``Multiplicative interactions and where to find them,''
\newblock in {\em International Conference on Learning Representations}, 2020.

\bibitem{Isola2017}
P.~Isola, J.~Zhu, T.~Zhou, and A.~A. Efros,
\newblock ``Image-to-image translation with conditional adversarial networks,''
\newblock in {\em Proceedings of the IEEE Conference on Computer Vision and
  Pattern Recognition}, July 2017.

\bibitem{Zhu2017}
J.~Zhu, R.~Zhang, D.~Pathak, T.~Darrell, A.~A Efros, O.~Wang, and E.~Shechtman,
\newblock ``Toward multimodal image-to-image translation,''
\newblock in {\em Advances in Neural Information Processing Systems}, 2017,
  vol.~30.

\bibitem{Mildenhall2021}
B.~Mildenhall, P.~P. Srinivasan, M.~Tancik, J.~T. Barron, R.~Ramamoorthi, and
  R.~Ng,
\newblock ``{NeRF}: Representing scenes as neural radiance fields for view
  synthesis,''
\newblock {\em Commun. ACM}, vol. 65, no. 1, pp. 99–106, dec 2021.

\bibitem{Shaham2021}
T.~R. Shaham, M.~Gharbi, R.~Zhang, E.~Shechtman, and T.~Michaeli,
\newblock ``Spatially-adaptive pixelwise networks for fast image translation,''
\newblock in {\em Proceedings of the IEEE/CVF Conference on Computer Vision and
  Pattern Recognition}, June 2021, pp. 14882--14891.

\bibitem{Martin2017}
M.~Arjovsky and L.~Bottou,
\newblock ``Towards principled methods for training generative adversarial
  networks,''
\newblock in {\em International Conference on Learning Representations}, 2017.

\bibitem{Menze2015}
B.~H. Menze, A.~Jakab, S.~Bauer, et~al.,
\newblock ``The multimodal brain tumor image segmentation benchmark
  ({BRATS}),''
\newblock {\em IEEE Transactions on Medical Imaging}, vol. 34, no. 10, pp.
  1993--2024, 2015.

\end{thebibliography}

\end{document}